\documentclass{icrc2009}

\usepackage{graphicx}   
\usepackage{caption}    
\usepackage[font=footnotesize]{subfig} 
\usepackage{fixltx2e}
\usepackage{url}

\newcommand{\shorttitle}[1]%
{\markboth{Proceedings of the 31\MakeLowercase{$^{st}$} ICRC, {\L}\'{o}d\'{z} 2009}{#1} }
\newcommand{\etal}{\MakeLowercase{\textit{et al. }}} 

\begin{document}
\title{Experimental search of bursts of very high energy gamma rays from primordial black holes}

\author{\IEEEauthorblockN{E.V. Bugaev\IEEEauthorrefmark{1}, V.B. Petkov\IEEEauthorrefmark{1}\IEEEauthorrefmark{2},
A.N. Gaponenko\IEEEauthorrefmark{1}, P.A. Klimai\IEEEauthorrefmark{1},
M.V. Andreev\IEEEauthorrefmark{1}\IEEEauthorrefmark{2}\IEEEauthorrefmark{3}, \\
I.M. Dzaparova\IEEEauthorrefmark{1}\IEEEauthorrefmark{2},
Zh. Sh. Guliev\IEEEauthorrefmark{1}, A.V. Sergeev\IEEEauthorrefmark{1}\IEEEauthorrefmark{2}\IEEEauthorrefmark{3},
V.I. Volchenko\IEEEauthorrefmark{1}, \\
G.V. Volchenko\IEEEauthorrefmark{1} and A. F. Yanin\IEEEauthorrefmark{1}}
                            \\
\IEEEauthorblockA{ \IEEEauthorrefmark{1}
 Institute for Nuclear Research, Russian Academy of Sciences, \\
 60th October Anniversary Prospect 7a, 117312 Moscow, Russia \\
 \IEEEauthorrefmark{2}
 Terskol Branch of the Institute of Astronomy of the Russian Academy of Sciences \\
 \IEEEauthorrefmark{3}
 International Center for Astronomical,  Medical and Ecological Research, \\
 National Academy of Sciences of Ukraine
 }
}

\shorttitle{Bugaev \etal Experimental search of bursts...}
\maketitle

\begin{abstract}

The technical procedure of a search of bursts of very high energy gamma rays from
evaporation of primordial black holes on air-shower array "Andyrchy" of Baksan
Neutrino Observatory of Institute for Nuclear Research is described. The theoretical
model used in the present work assumes that the chromosphere around the evaporating
black hole does not form. For minimization of the cosmic ray background the method
of multidimensional analysis of modelled as well as experimentally detected events
is applied. The new upper limit on the concentration of evaporating primordial
black holes in the local region of Galaxy is obtained. The comparison of the results
of different experiments is given.

\end{abstract}


\section{Introduction}

Primordial black holes (PBHs) can be formed in the early Universe through the gravitational collapse
of primeval cosmological density fluctuations. Therefore, the formation probability of PBHs and
their observational manifestations depend significantly on how the primeval density fluctuations
emerged and developed. Theoretical predictions of the PBH formation probability depend
strongly on the adopted theory of gravitation and on the model of
gravitational collapse. The evaporation of black holes \cite{Hawking:1974sw}, on which their
experimental search is based, has not
been completely studied either. Thus, PBH detection will provide valuable information about the early
Universe and can be a unique test of the general theory of relativity, cosmology, and quantum gravity
\cite{Carr:2003bj}.
Direct searches for the bursts of gamma rays from the evaporations of PBHs have been carried out in several works during the last 15 years \cite{Alexandreas:1993zx, Amenomori, Funk, Connaughton:1998ac, Linton:2006yu, Petkov:2008rz}.

The search for high energy gamma ray bursts on a shower array reduces to the search of space and time
correlations (clusters) of registered extensive air showers (EAS). If time intervals used for
the search are rather small, this search is performed in a horizontal reference frame. For each
shower $i$ having absolute registration time $t_i$ and arrival angles $(\theta, \phi)_i$ the
cluster of events $i, i+1, i+2, ..., i+n-1$ is sought for, using a condition that arrival directions
should differ less than $\alpha_r$ from the weighted mean direction. Thus, each cluster is
characterized by multiplicity $n$, duration $\Delta t$, absolute time $T$, and arrival direction
$(\theta, \phi)$.

During such a search, experimentally obtained dependencies (e.g., cluster registration frequencies for
each $n$) are compared to the ones expected from the background of accidental coincidences. If measured
frequencies of cluster registration can be explained with distributions expected for accidental coincidences,
one can obtain the constraints on a gamma ray burst frequency producing clusters with a given multiplicity
(and thus having particular energy flux).

Evaporating PBHs which have almost reached their final evaporation state are a possible source of high energy
gamma ray bursts. Since the calculated temporal and energy characteristics of such bursts
depend on the theoretical evaporation model \cite{Bugaev:2007py}, the technique of an experimental search
and the constraints imposed on the PBH number density in the local Universe are model dependent.
PBHs can be sought for in experiments on arrays designed to detect
EASs from cosmic rays with effective primary gamma-ray energies of 10 TeV or higher only within
the assumption that the 
evaporation model without a chromosphere \cite{MacGibbon:1990zk} is correct
(in such a model, evaporated particles do not interact with each other).
The duration of the high-energy GRBs predicted by chromospheric evaporation models is too short, much
shorter than the dead time of EAS arrays, and other methods have to be applied for
the PBHs search in this case \cite{Petkov:2008rv}.

It should be noted that the duration of the high energy GRBs is fairly short in the evaporation model
without a chromosphere as well. Therefore, the effect of the array dead time on the burst detection probability
should be taken into account when interpreting the experimental data from EAS arrays with a high
threshold energy of the primary photons. Such analysis have been performed in our
previous work \cite{Petkov:2008rz}.

In the experiments searching for gamma ray bursts produced
by PBHs, the constraints on the PBH number density are usually obtained from a condition of absence
of clusters with a given multiplicity and duration. Thus, an one-dimensional task is solved - each cluster
(not depending on multiplicity) corresponds to the point with coordinate $\Delta t$ on a time axis,
and cluster duration distribution is analyzed then. However, for the cluster with $n$ showers,
the number of independent parameters (time intervals) is $n-1$ and such a cluster can be represented as a
point in space with $n-1$ dimensions. One can expect that in those cases when time characteristics
of PBH produced clusters are different from ones arising due to accidental coincidences, the increase
the number of parameters can help to separate PBH events from background ones. In this case, due
to background reduction, the constraint on the PBH number density obtained on a given array
can be improved.

This paper is devoted to exploring this idea and improving the constraint on
PBH number density obtained previously on Andyrchy array \cite{Petkov:2008rz}.
The detailed description of this experimental facility can be found in \cite{Petkov:2006zz}.

\section{The experiment}

\begin{figure}[!t]
\centering
\includegraphics[width=1.03 \columnwidth ]{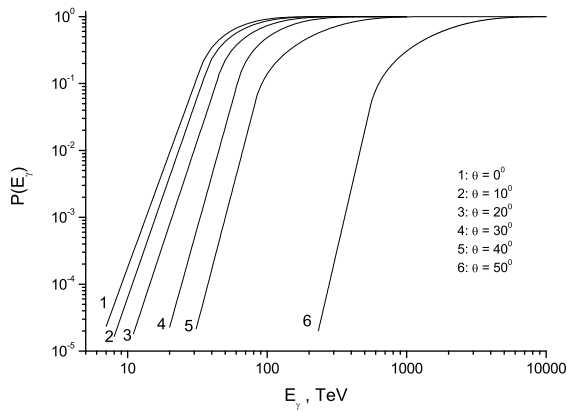}
\caption{Probabilities of EAS detection by the Andyrchy array
versus primary gamma-ray photon energy for various zenith angles. }
\label{fig2}
\end{figure}

The detection probabilities $P(E_\gamma, \theta)$ of the EASs
generated by primary gamma-rays with energy $E_\gamma$ falling on Andyrchy array at zenith angle
$\theta$ were determined by simulating of electromagnetic cascades in the atmosphere and the detector
(see Fig. \ref{fig2}) \cite{Smirnov2005}. Total number of gamma rays which can be 
detected by the array (integral burst profile) is given by
\begin{equation}
N_\gamma(\theta, t_l) = \int\limits_0^\infty dE_\gamma P(E_\gamma, \theta) dN_\gamma/dE_\gamma.
\end{equation}
It depends on the time $t_l$ left until the end of PBH evaporation.

In our previous work \cite{Petkov:2008rz} we have defined the burst duration $t_b$ for the given array
as time until the end of PBH evaporation during which 99\% of photons which can be detected by this array
are evaporated. This interval was used as the interval for search of clusters from PBHs, i.e., $T_s=t_b$
(and shower clusters with $\Delta t \le t_b$ were sought for).

\begin{figure}
\centering
\includegraphics[width=1.03 \columnwidth ]{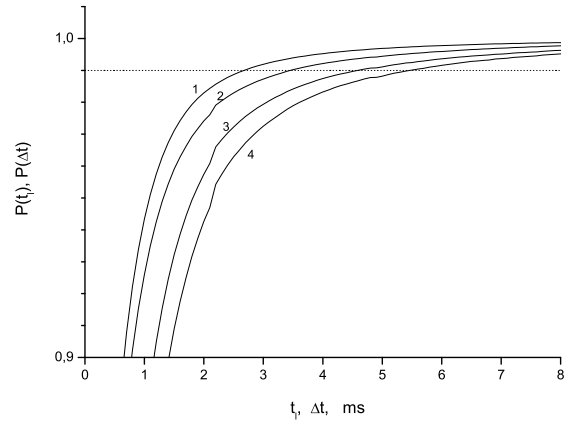}
\caption{Curves 2,3,4: probabilities $P(\Delta t)$ for the clusters with $n=2,3,4$ to have the duration
in the time interval between 0 and $\Delta t$, for zenith angle $\theta=30^\circ$. Curve 1 shows,
for comparison, the probability $P(t_l)$ for photons to be evaporated during time interval between 0 and $t_l$.}
\label{fig7}
\end{figure}

\begin{figure}[!b]
\centering
\includegraphics[width=1.03 \columnwidth ]{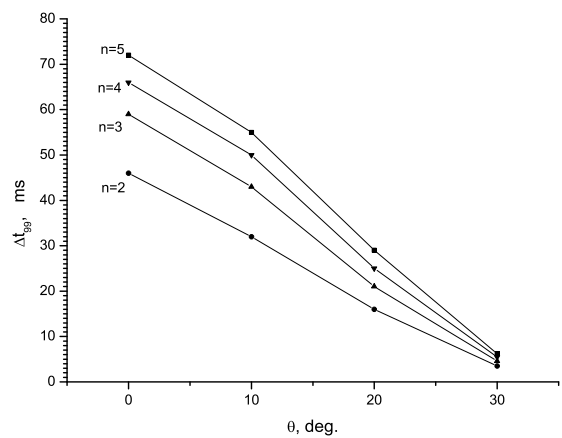}
\caption{The dependencies of  $\Delta t_{99}(n)$ on zenith angle for cluster
multiplicity $n=2,3,4,5$.}
\label{fig8}
\end{figure}

Because the exact moment of PBH evaporation is unknown, the obtained experimental distributions for
cluster lengths should be compared not with the burst profile, but with the calculated distribution of
clusters with given multiplicity. Fig. \ref{fig7} shows, for $\theta=30^\circ$, the calculated integral
$\Delta t$ distributions for clusters with $n=2,3,4$ (curves 2,3 and 4).
In the calculation, the dead time of the array had been taken into account.
In the same Fig. \ref{fig7} the burst profile (i.e., the
probability to detect a gamma particle from evaporating PBH in the interval of time between $t=t_l$
and full evaporation) is also shown (curve 1). It is seen from this figure that 99\%-level for clusters is achieved
at durations much larger than $t_b$.

Correspondingly, the time interval for search $T_s$ should be taken now as $T_s=\Delta t_{99}(n)$ which
we define as the time interval in which 99\% of clusters with multiplicity $n$ fall. Fig. \ref{fig8}
shows the dependence of $\Delta t_{99}(n)$ on zenith angle for different $n$. One can note
that time intervals $\Delta t_{99}(n)$ shown in this figure are larger than $t_b$ (e.g., for
$\theta=0^\circ$, $t_b=40$ ms - see \cite{Petkov:2008rz}). Because of this, maximum multiplicities
for registered clusters also increase: for the case $T_s=t_b$, maximum multiplicities were 4,4,3,2
for zenith angles $\theta=0^\circ,10^\circ,20^\circ,30^\circ$; for the case $T_s=\Delta t_{99}(n)$ these
values are 4,4,4,3 for the same corresponding values of zenith angles.

\begin{figure}[!t]
\centering
\includegraphics[width=1.03 \columnwidth ]{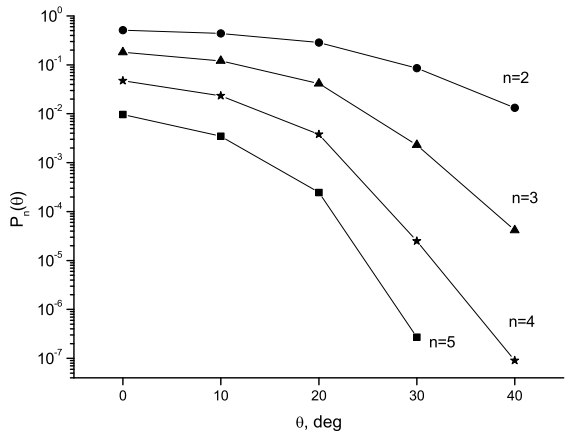}
\caption{Probabilities of registering the gamma ray clusters from
the evaporating PBH, for the Andyrchy array, as a function of zenith
angle.}
\label{fig9}
\end{figure}

The probability to detect all showers hitting the array in a cluster with multiplicity $n$ was
calculated using Monte-Carlo method taking into account the dead time of the array (only those
clusters were taken for which time interval between consecutive events was larger than a dead
time of the array $t_d\sim 1$ ms), see Fig. \ref{fig9}.

\begin{figure}
\centering
\includegraphics[width=1.03 \columnwidth ]{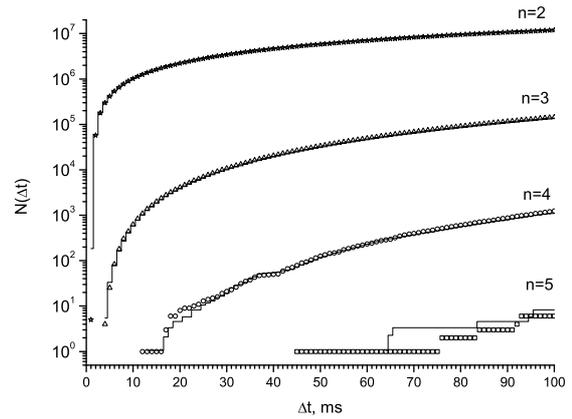}
\caption{Integral duration distributions for clusters with multiplicity $n=2,3,4,5$, detected by
the array for the whole region of zenith angles. Points represent the experimental results; lines
show the calculated distributions expected from the cosmic ray background.}
\label{fig10}
\end{figure}

The result of the search for shower clusters from evaporating PBHs is the set of duration distributions
of clusters for different $n$ and $\theta$. Fig. \ref{fig10} shows the experimentally measured
distributions for a full observational time $\approx 1100$ days, summarized over all zenith angles. For the
search of clusters, we took $\alpha_r=7.0^\circ$ - such a region contains 90\% of events
from a point source. In the same figure, we show modeled distributions for clusters produced by accidental
coincidences of regular EASs.

\begin{figure}[!b]
\centering
\includegraphics[width=11 cm, trim=80 0 0 0]{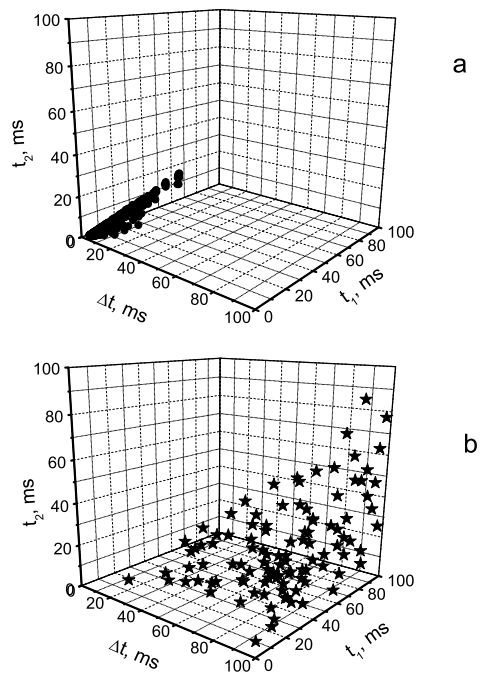}
\caption{3-dimensional distributions for clusters with $n=4$, zenith angle is $\theta=30^\circ$.
a) Events, expected from the evaporating PBH; b) events registered by the Andyrchy array.}
\label{fig11}
\end{figure}

It is seen from Fig. \ref{fig10} that, generally, experimentally registered cluster distributions
can be explained as accidental coincidences of EASs produced by cosmic rays.

\section{Multidimensional analysis}
For clusters with $n$ showers, as we have seen, number of independent parameters (time intervals) is $n-1$,
and such a cluster can be represented as a point in space with corresponding number of dimensions.
As an example, Fig. \ref{fig11} shows 3-dimensional distributions of clusters expected from PBHs (Fig. \ref{fig11}-a)
and experimentally registered clusters (Fig. \ref{fig11}-b) for the case of $n=4$ and $\theta=30^\circ$.
It is seen that the measured clusters and the ones expected from PBHs occupy different regions in 3-dimensional
space of time intervals.

In the subsequent analysis, the probabilities for the modeled PBH event to get into a $(n-1)$-dimensional
cell with size 1 ms in each dimension were calculated for each range of zenith angles $\theta$ and
cluster multiplicities $n$. These probabilities were summarized for each cell which contained
experimentally registered events. If this sum did not exceed $10^{-3}$, we assumed, for PBH
number density constraint calculation, that number of registered PBH events is zero
(for particular values of $n$ and range of zenith angles). We have obtained $3,3,3,$ and $2$ for
maximum cluster multiplicities of registered events, for $\theta=0^\circ,10^\circ,20^\circ,30^\circ$,
correspondingly.

\section{Results and conclusions}

The constraints on the number density of evaporating PBHs in the local region of Galaxy were
obtained using technique described in work \cite{Linton:2006yu}. Let a PBH be located at
distance $r$ from the array and be seen from it at zenith angle $\theta$. The mean number of gamma-ray photons
detected by the array over the burst duration is then
\begin{equation}
\bar n(\theta) = \frac{\epsilon N(t_b(\theta))S(\theta)}{4\pi r^2}\; ,
\end{equation}
where $S(\theta)$ is the array area and $\epsilon = 0.9$ is the fraction
of the events from a point source that fell into an angular cell with a size of $\alpha_r$.
The number of bursts detected over the total observation time T can be represented as
\begin{equation}
N=\rho_{\rm pbh} T V_{\rm eff} \; ,
\end{equation}
with effective volume of the space surveyed by the array
\begin{equation}
V_{\rm eff}=\int d \Omega {\int \limits_{0}^{\infty} dr
r^2 F(n(\theta),\bar n(\theta))}.
\end{equation}
Here, $F(n,\bar n)$ is the detection probability of a cluster of $n$ or more EASs with the mean value of $\bar n$.
It can be expressed through the Poisson probabilities for having $i$ showers (and $i=n,n+1,...$) with average value of
$\bar n$  and probabilities $P(i(\theta), t_d)$ [see Fig. \ref{fig9}] to detect all $i$ showers
having a dead time for one event $t_d$:
\begin{equation}
F(n,\bar n) = \sum_{i=n}^{\infty} P(i(\theta),t_d)\cdot  \frac{e^{-\bar n} {\bar n}^i}{i!}.
\end{equation}
For the calculation of the effective volume, we take $n(\theta) =
n'(\theta) + 1$ (this means that the
distributions of the detected clusters in multiplicity can be explained by the
background of accidental coincidences; $n'(\theta)$ is the maximal multiplicity of the detected cluster 
obtained using multidimensional analysis).


Numerically, we obtain $V_{\rm eff} = 2.8\times 10^{-9}$ pc$^{3}$. If the evaporating PBHs are distributed
uniformly in the local region of the Galaxy, then the upper limit $\rho_{\rm lim}$ on the number density
of evaporating PBHs at the 99\% confidence level can be calculated from the formula
\begin{equation}
\rho_{\rm lim} = \frac{4.6}{V_{\rm eff}\cdot T}.
\end{equation}
As a final result, we obtain $\rho_{\rm lim} = 5.4 \times 10^8$ pc$^{-3}$yr$^{-1}$.
This is slightly better than the result obtained in our previous work \cite{Petkov:2008rz}
with the same set of data but without using
multidimensional analysis
($\rho_{\rm lim} = 8.2 \times 10^8$ pc$^{-3}$yr$^{-1}$).

The detailed comparison of the results of different experiments searching for evaporating PBHs can be
found in \cite{Petkov:2008rz}. This work slightly improves the PBH number density constraint of
\cite{Petkov:2008rz}, but it is still significantly weaker than the best (to date) limit
obtained in the experiment on the Whipple Cherenkov telescope \cite{Linton:2006yu}
($\rho_{\rm lim} = 1.08 \times 10^6$ pc$^{-3}$yr$^{-1}$). However, it should be noted that the
effective gamma-ray energy in our experiment is two orders of magnitude higher than that of the
Whipple telescope. Thus, our upper limit pertains not to black holes in general, but to black holes with certain
properties (those emitting 100-TeV gamma-rays at the end of their evaporation during bursts lasting
$\sim 10$ ms).

\bigskip

{\bf Acknowledgements.} This work was supported by the the Russian Foundation for Basic Research
(Grants No. 06-02-16135, 08-07-90400 and 09-02-90900). This work was also supported in part by the
"Neutrino Physics and Astrophysics" Program for Basic Research of the Presidium of the
Russian Academy of Sciences and by "State Program for Support of Leading Scientific Schools"
(Project No. NSh-321.2008.2).


\begin{thebibliography}{99}

\bibitem{Hawking:1974sw}
  S.~W.~Hawking,
  Commun.\ Math.\ Phys.\  {\bf 43}, 199 (1975)
  [Erratum-ibid.\  {\bf 46}, 206 (1976)].

\bibitem{Carr:2003bj}
  B.~J.~Carr,
  Lect.\ Notes Phys.\  {\bf 631}, 301 (2003)
  [arXiv:astro-ph/0310838].

\bibitem{Alexandreas:1993zx}
  D.~E.~Alexandreas {\it et al.},
  Phys.\ Rev.\ Lett.\  {\bf 71}, 2524 (1993).

\bibitem{Amenomori} M.Amenomori \etal, Proc. 24th International Cosmic Ray Conference, Rome (Italy), v.2, 112 (1995).

\bibitem{Funk} B.Funk \etal, Proc. 24th International Cosmic Ray Conference, Rome (Italy), v.2, 104 (1995).

\bibitem{Connaughton:1998ac}
  V.~Connaughton {\it et al.},
  Astropart.\ Phys.\  {\bf 8}, 179 (1998).

\bibitem{Linton:2006yu}
  E.~T.~Linton {\it et al.},
  JCAP {\bf 0601}, 013 (2006).

\bibitem{Petkov:2008rz}
  V.~B.~Petkov {\it et al.},
  Astron.\ Lett.\  {\bf 34}, 509 (2008)
  [Pisma Astron.\ Zh.\  {\bf 34}, 563 (2008)]
  [arXiv:0808.3093 [astro-ph]].

\bibitem{Bugaev:2007py}
  E.~Bugaev, P.~Klimai and V.~Petkov,
  Proc. 30th International Cosmic Ray Conference, Mexico , v.3, 1123 (2007).
  arXiv:0706.3778 [astro-ph].

\bibitem{MacGibbon:1990zk}
  J.~H.~MacGibbon and B.~R.~Webber,
  Phys.\ Rev.\  D {\bf 41}, 3052 (1990).

\bibitem{Petkov:2008rv}
  V.~B.~Petkov, E.~V.~Bugaev, P.~A.~Klimai and D.~V.~Smirnov,
  JETP Lett.\  {\bf 87}, 1 (2008)
  [Pisma Zh.\ Eksp.\ Teor.\ Fiz.\  {\bf 87}, 3 (2008)]
  [arXiv:0803.2313 [astro-ph]].

\bibitem{Petkov:2006zz}
  V.~B.~Petkov {\it et al.},
  Instrum.\ Exp.\ Tech.\  {\bf 49}, 785 (2006)
  [Prib.\ Tekh.\ Eksp.\  {\bf 49}, 50 (2006)].

\bibitem{Smirnov2005}
  D.~V.~Smirnov \etal,
  Bull. Rus. Acad. Sci.: Physics, {\bf 69} N3: 413 (2005).


\end{thebibliography}
\end{document}